\documentclass[a4paper]{article}
\usepackage{listings}
\usepackage{xcolor}
\usepackage{multirow}
\usepackage{jheppub}  
\usepackage{dcolumn}
\usepackage[english]{babel}
\usepackage[utf8x]{inputenc}
\usepackage[T1]{fontenc}
\usepackage{palatino}
\usepackage{amsmath}
\usepackage{afterpage}
\usepackage{graphicx, subcaption}
\usepackage[colorinlistoftodos]{todonotes}
\usepackage[colorlinks=true]{hyperref}
\usepackage{makeidx}
\newcommand{\be}{\begin{equation}}  
\newcommand{\ee}{\end{equation}}  
\newcommand{\bea}{\begin{eqnarray}}  
\newcommand{\eea}{\end{eqnarray}}

\newcommand{\media}[1]{\left\langle #1 \right\rangle}

\hypersetup{colorlinks=true,citecolor=blue,urlcolor=blue,linkcolor=blue}

\makeindex
\begin{document}

\vspace*{1.2cm}

\thispagestyle{empty}
\begin{center}

{\LARGE \bf Scaled transverse-momentum spectra as a probe of collective dynamics in heavy-ion collisions}

\par\vspace*{7mm}\par

{

\bigskip

\large \bf Thiago Siqueira Domingues and Matthew Luzum for the ExTrEMe Collaboration}

\bigskip

{\large \bf  E-Mail: thiago.siqueira.domingues@usp.br}

\bigskip

{University of São Paulo - SP}

\bigskip

{\it Presented at the Workshop of Advances in QCD at the LHC and the EIC, CBPF, Rio de Janeiro, Brazil, November 9-15, 2025}

\vspace*{15mm}

\end{center}
\vspace*{1mm}
\begin{abstract}
-We investigate a scaling property of transverse-momentum spectra in ultrarelativistic heavy-ion collisions obtained by removing the global scales of multiplicity and mean transverse momentum. The resulting dimensionless observable isolates the intrinsic shape of the spectrum and reveals an approximate universality across collision centralities, systems, and energies. Hydrodynamic simulations reproduce this scaling on an event-by-event basis, indicating that it may originate from the collective dynamics of the quark–gluon plasma. 

Using Gaussian-process emulators trained on the JETSCAPE hybrid model, we perform a Bayesian analysis incorporating the scaled spectra as observables. The results demonstrate that the spectral shape provides independent constraints on key properties of the medium, including pre-equilibrium dynamics and initial-state granularity, while exposing tensions with parameter regions preferred by traditional $p_T$-integrated observables. 

We further explore an analogous scaling of transverse-mass spectra and observe a comparable universality across centralities and hadron species. These results suggest that scaled spectra provide a powerful new probe of collective dynamics and offer complementary constraints for the quantitative characterization of QCD matter created in heavy-ion collisions.
\end{abstract}
\section{Introduction}
\label{sec:intro}
Ultrarelativistic nuclear collisions create the Quark–Gluon Plasma (QGP), a nearly perfect fluid of deconfined quarks and gluons formed in the hot and dense environment produced at RHIC and the LHC~\cite{Busza:2018rrf}. Over the past two decades, a wealth of experimental evidence has demonstrated that the QGP exhibits strong collective behavior. Anisotropic flow observables, quantified through the Fourier coefficients $v_n$, have emerged as the primary probes of this collectivity~\cite{JETSCAPE:2020mzn, Gardim:2012yp, Luzum:2013yya, ALICE:2016kpq}. These observables encode the response of the system to initial-state geometry and fluctuations, and their successful description by relativistic hydrodynamics has established the fluid paradigm as the standard framework for interpreting heavy-ion data. One of the most important observables in these extreme environments is the transverse momentum distribution of charged and identified particles. Despite this progress, the information contained in the full transverse momentum distribution, $dN/dp_T$, remains less thoroughly explored~\cite{ExTrEMe:2024fxt}.

Traditionally, single-particle spectra have been used to extract effective temperatures, radial flow parameters, and kinetic freeze-out conditions through blast-wave fits~\cite{Tomasik:2024uuq}. However, such extractions rely on model-dependent assumptions and typically focus on integrated or averaged features rather than the event-by-event structure or the underlying response of spectra to dynamical fluctuations. In this study, we aim to investigate the reduced scaled observable proposed by the Extreme Collaboration~\cite{ExTrEMe:2024fxt, Domingues:2025tkp, Domingues:2026mis},
designed to isolate the intrinsic shape of the transverse momentum spectrum, 
\begin{equation}
U(x_T) \equiv \frac{\media{p_T}}{N}\frac{dN}{dp_T}=\frac{1}{N}\frac{dN}{dx_T}, 
\label{eq:universal}
\end{equation}
with $x_T=p_T/\media{p_T}$. By normalizing by the multiplicity $N$ and the event-averaged mean transverse momentum 
$\media{p_T}$, this construction removes the dominant global scales that vary with centrality, system size, and energy~\cite{ExTrEMe:2024fxt}. What remains is a dimensionless function of the scaled transverse momentum $x_T$ that captures the intrinsic spectral shape. Remarkably, the scaled spectra collapse onto a nearly universal curve, even event-by-event, across centralities (see Fig.\ref{fig:pt_spectra_all}) and systems (Pb–Pb, Xe–Xe, and p–Pb), breaking only at high $p_T$ and in very small systems (p–p)~\cite{ExTrEMe:2024fxt}. Hydrodynamic models reproduce the scaling event-by-event, establishing it as a possible signature of collective fluid dynamics.
In such an event-by-event analysis, this observable becomes related to the so-called $p_T$-differential radial flow coefficient $v_0(p_T)$~\cite{Schenke:2020uqq, Parida:2024ckk}, with the double scaling $v_0(p_T)/v_0$ plotted against $x_T$, signaling that this coefficient presents nearly the same universal behavior across centrality, energy, and systems.

While Bayesian analyzes have so far relied mainly on $p_T$-integrated observables such as $dN/dy$, $\langle p_T \rangle$, and $v_n$~\cite{JETSCAPE:2020mzn}, the strong fluid connection of this scaling makes the $p_T$ spectra a natural ingredient that could provide independent constraints on QCD transport properties. As the onset and breakdown of hydrodynamic behavior remain open problems~\cite{Domingues:2024pom}, the deviations of the nearly universal scaling offer a powerful tool to pinpoint the limits of collectivity in hadronic collisions. In the following sections, we explore the properties of $U(x_T)$, quantify its degree of universality across different experimental conditions, and compare the observed scaling to predictions from state-of-the-art hydrodynamic models. We further discuss its implications for understanding radial flow fluctuations, particle production mechanisms, and the emergence of fluid-like behavior in small collision systems. This proceedings is organized as follows: in Sec.~\ref{sec:universality_over_params}, we present a model-to-data analysis of the scaled spectra $U(x_T)$ using the JETSCAPE state-of-the-art model to compare with ALICE data. In the same section, we discuss in Sec.~\ref{subsec:gsa} a global sensitivity analysis performed over the whole parameter space to determine the most important model parameters. In Sec.~\ref{sec:universality_breakdown} we discuss deviations of the nearly universal behavior across the parameter space pointing out the most influential physical effects for this universality breakdown. The scaling for the transverse mass spectra is displayed in Sec.~\ref{sec:transverse_mass_spectra}, where we present a universality over centrality and discuss a potential universality over different hadronic species. Finally, in Sec.~\ref{sec:conclusions} we design our conclusions and future perspectives.
\section{Incorporating scaled spectra into Bayesian analysis}
\label{sec:universality_over_params}
In this section, we present a Bayesian analysis of the universal scaling properties of the transverse momentum spectra in Pb--Pb collisions at $\sqrt{s_{NN}} = 2.76~\mathrm{TeV}$. Our work employs the simulation data of the JETSCAPE collaboration presented in Ref.~\cite{JETSCAPE:2020mzn}, which was used to train Gaussian--process (GP) emulators to investigate how well state-of-the-art hydrodynamic modeling reproduces the experimentally observed universality of the scaled pion spectra. See~\cite{Domingues:2026mis} for further details.
\subsection{Comparison to Experimental Data}
\label{subsec:model_to_data}
\begin{figure}
    \centering
    \includegraphics[width=\linewidth]{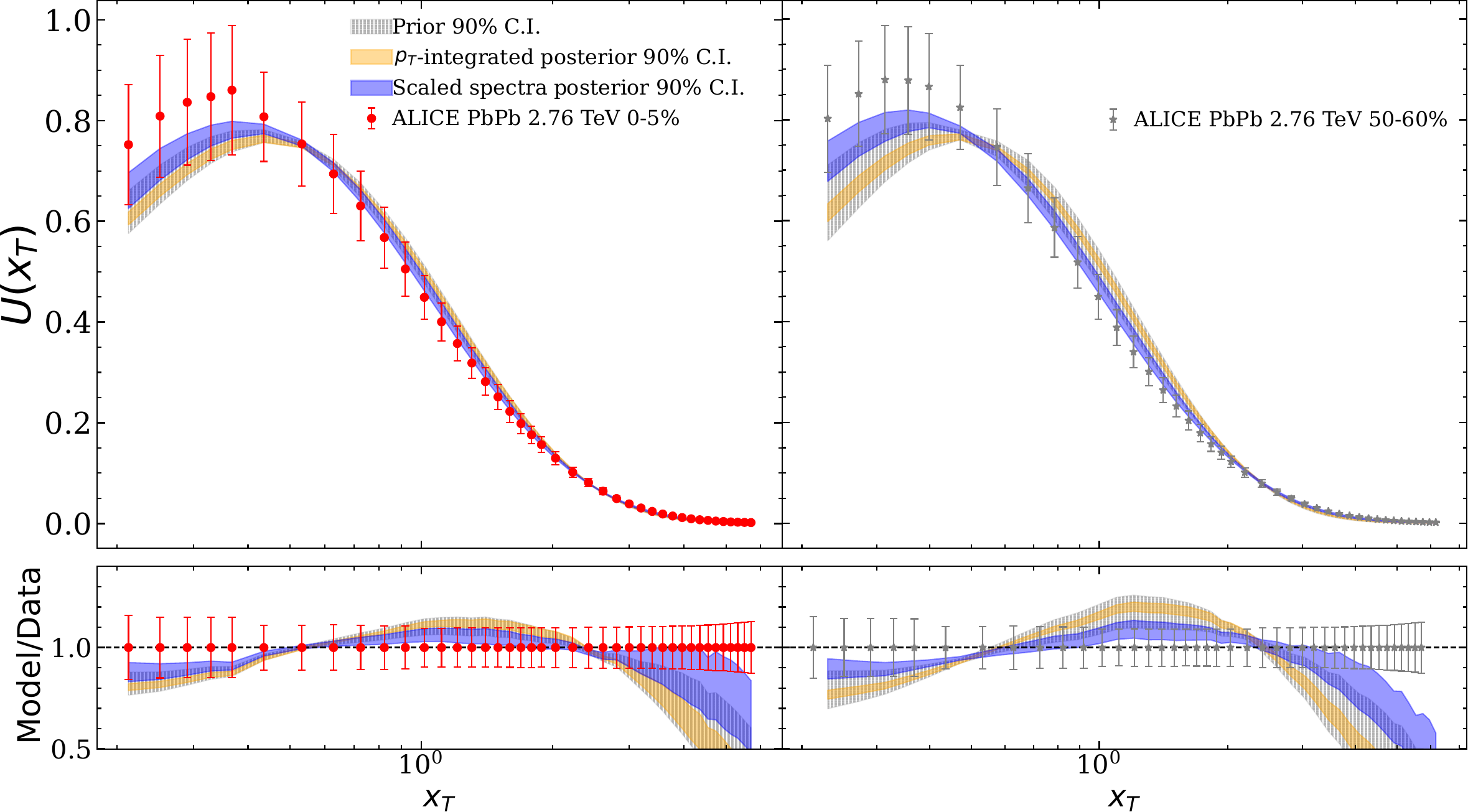}
    \caption{Prior (hatched gray bands) and posterior (orange and blue bands) model predictions for the scaled spectra $U(x_T)$ obtained with the Grad viscous correction, compared with ALICE Pb--Pb data at $\sqrt{s_{NN}} = 2.76~\mathrm{TeV}$~\cite{ALICE:2013mez} for centralities 0--5\% and 50--60\%. The narrow prior bands reflect the intrinsic universality of $U(x_T)$, while the posterior calibration sharpens these predictions, capturing the overall shape but revealing systematic deviations at low and high $x_T$. Experimental data points include combined uncorrelated statistical and systematic uncertainties. The original JETSCAPE posterior calibrated to $p_T$-integrated observables~\cite{JETSCAPE:2020mzn} are displayed by the orange bands. The scaled-spectra posterior, represented by the blue bands, is consistent with data but differs systematically from the integrated-observable posterior, indicating a tension between the two calibration strategies and highlighting the sensitivity of $U(x_T)$ to distinct regions of parameter space.}
    \label{fig:model_vs_data_Grad}
\end{figure}

Figure~\ref{fig:model_vs_data_Grad} illustrates the comparison between experimental ALICE data~\cite{ALICE:2013mez} and model predictions obtained from the prior and two different posterior distributions. The prior predictions produce surprisingly narrow 90\% credible intervals, indicating that the universality of $U(x_T)$ is already an intrinsic feature of the hydrodynamic model.
After calibration to the scaled spectra, the posterior bands sharpen and generally reproduce the data. However, a characteristic pattern appears: 
(i) mild underestimation at low $x_T$, 
(ii) overshoot at intermediate $x_T$, and 
(iii) underestimation in the semi-hard region. 
These deviations reflect the transition from soft hydrodynamic physics to the regime where non-equilibrium and semi-hard contributions become non-negligible ($p_T \gtrsim 1.5$–2 GeV), a region where hydrodynamics is not expected to be fully reliable.
When compared with the original JETSCAPE calibration~\cite{JETSCAPE:2020mzn} (based solely on $p_T$-integrated observables), we find a noticeable tension: the two posterior distributions prefer distinct regions of parameter space. This indicates that the scaled spectra are sensitive to different physics than traditional observables such as $dN/dy$, $\langle p_T \rangle$, $\delta p_T / \langle p_T \rangle$, and the $p_T$-integrated flow harmonics $v_n\{2\}$.
\subsection{Global Sensitivity Analysis}
\label{subsec:gsa}
To identify which parameters govern the shape of $U(x_T)$, we perform a global sensitivity analysis (GSA) using Sobol indices computed with the GP emulators. Four parameters dominate:
\begin{itemize}
    \item the free-streaming timescale $\tau_R$, which controls pre-equilibrium evolution;
    \item the maximum of the bulk viscosity $(\zeta/s)_{\max}$, influencing entropy production and radial flow;
    \item the temperature at which the bulk viscosity peaks, $T_\zeta$;
    \item the Gaussian nucleon width $w$, which determines the granularity of the initial state.
\end{itemize}
These parameters collectively determine the curvature, steepness, and centrality dependence of the scaled spectra. A comparison of posterior distributions reveals strong tension between observables: the scaled-spectra calibration favors small nucleon widths, $w \sim 0.6$ fm, corresponding to granular initial conditions, whereas the $p_T$-integrated calibration prefers smoother profiles with $w \sim 1.0$ fm. This behavior aligns with other observables that push toward small $w$, such as the total nucleon--nucleon cross section~\cite{Nijs:2022rme} and the $\rho_2$ correlation between $\langle p_T\rangle$ and $v_2$~\cite{Giacalone:2021clp}. Again, we refer the readers to Ref.~\cite{Domingues:2026mis} for further details.
\section{Universality breakdown across the parameter space}
\label{sec:universality_breakdown}
While the scaled spectra $U(x_T)$ exhibit a remarkable degree of universality across centrality classes, a closer analysis reveals that this scaling is not exact and can break down for certain combinations of model parameters. To identify where in the multidimensional parameter space this universality is violated and which physical quantities control such deviations, we developed a systematic quantitative approach combining Gaussian process emulation with GSA.

\begin{figure}
  \centering
  \includegraphics[width=\linewidth]{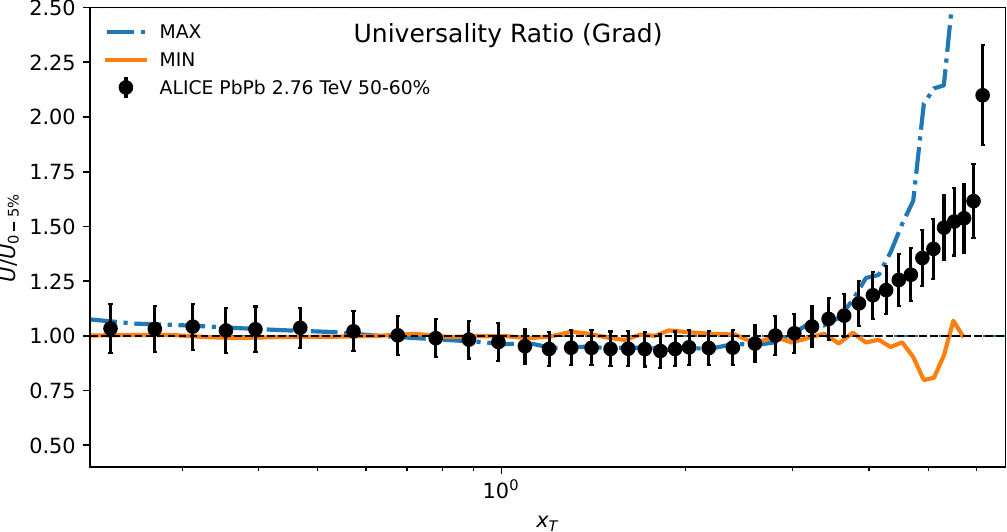}
  \caption{Deviation from universal scaling for the scaled spectra $U(x_T)$ comparing the most central (0--5\%) and most peripheral (50--60\%) events. The curves labeled ``Max" and ``Min" represent the envelope of deviations obtained across the full model parameter space using the Grad viscous correction. The model predictions are compared with experimental measurements from ALICE~\cite{ALICE:2013mez}.}
  \label{fig:universality_breakdown}
\end{figure}

The first step consists of defining a measure of universality breakdown across centralities. For each design point in the parameter space, we computed the model-predicted scaled spectra $U(x_T)$ for all centrality classes. The deviation from perfect universality was quantified relative to the most central class (0–5\%) by taking pairwise differences in $U(x_T)$ between each centrality and the reference. We emphasize that using differences rather than ratios provides better numerical stability and physical interpretability. Ratios can suffer from singularities when the denominator becomes small and from discontinuities at very low or high $x_T$, while differences remain smooth and well-defined across the entire momentum range. In Fig.~\ref{fig:universality_breakdown} we display the universality ratio comparing the most central (0-5\%) and the most peripheral (50-60\%) centrality bins. As shown, the ``Max" curve represents the maximum deviation over the parameter space and better describes the ALICE experimental data, especially in larger $x_T$ regions where we expect larger deviations.

\begin{figure}
  \centering
  \includegraphics[width=\linewidth]{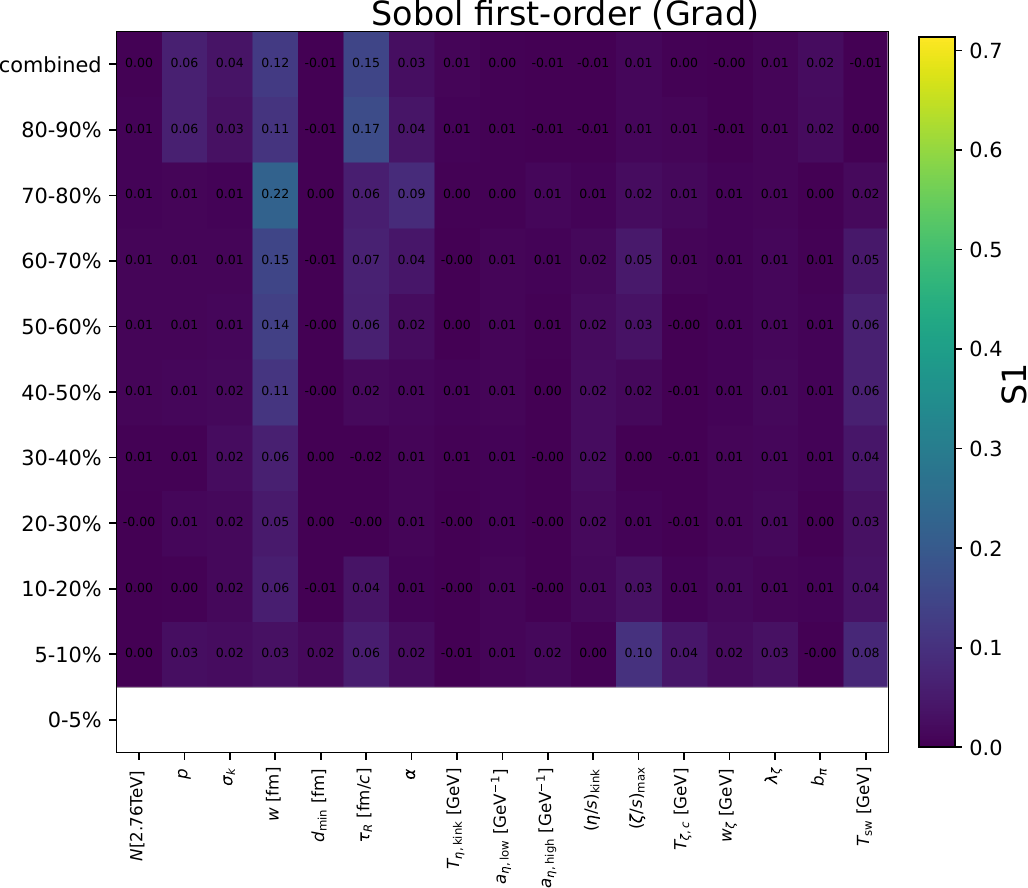}
  \caption{First-order Sobol indices quantifying the contribution of each model parameter to deviations from universality in $U(x_T)$ for different centrality classes in Pb--Pb collisions at $\sqrt{s_{NN}} = 2.76~\mathrm{TeV}$. The analysis uses GP emulators trained on JETSCAPE simulations with the Grad viscous correction. The nucleon width $w$ dominates the breakdown of universality in peripheral collisions, while the free-streaming time $\tau_R$ and maximum bulk viscosity $(\zeta/s)_{\rm max}$ become more important in mid-central events.}
  \label{fig:sobol_index_breakdown}
\end{figure}

For each design point, we then evaluated the root-mean-square error (RMSE) of these differences across all 41 $x_T$ bins, yielding a single scalar quantity that summarizes the overall magnitude of universality breaking for each centrality. A combined scalar was obtained by taking the RMS over all centralities, representing the total degree of non-universality for a given parameter set. The scalar quantities were subsequently emulated using GP regressors trained on the design points, allowing continuous interpolation over the full 17-dimensional parameter space. Finally, we applied GSA by computing the first-order Sobol indices for each model parameter with respect to this universality-deviation measure. These Sobol indices quantify the fraction of the total variance in the universality breakdown that is attributable to variations in each parameter individually. The Sobol first-order indices reveal clear trends in how universality breakdown emerges across the model parameter space. The results are summarized in Fig.~\ref{fig:sobol_index_breakdown}.
For the most peripheral collisions, the Gaussian nucleon width $w$ is the dominant contributor to the observed non-universality, indicating that variations in the effective nucleon size primarily control deviations from the universal scaled shape in small or dilute systems. Smaller nucleon widths correspond to more granular initial conditions and hence larger event-by-event fluctuations, which tend to destroy the near-universal scaling observed in the bulk-dominated central events. The dependence on centrality is not uniform. While $w$ plays a leading role in the periphery, other parameters such as the free-streaming relaxation time $\tau_R$ and the maximum bulk viscosity $(\zeta/s)_{\rm max}$ contribute more significantly for mid-central collisions, where the transition between strongly and weakly collective regimes occurs. This variation across centralities indicates that different stages of the QGP evolution become relevant to the universality properties in different geometrical and density conditions.

The combined universality breakdown measure, obtained by the RMS across all centralities, confirms that the nucleon width $w$ and the free-streaming time $\tau_R$ are the two most influential parameters overall.
These results provide strong evidence that the breakdown of universality in $U(x_T)$ is primarily driven by initial-state granularity and pre-equilibrium dynamics rather than by uncertainties in the viscous transport coefficients. In physical terms, these findings suggest that the universality of scaled particle spectra reflects a delicate balance between the smoothing effects of collective expansion and the microscopic granularity imprinted by the initial conditions.
When the effective nucleon width $w$ is small, the resulting lumpy initial profiles enhance local flow gradients and generate deviations from the universal spectral scaling, particularly in peripheral events where the system size is smaller and fluctuations dominate. In contrast, larger $w$ values and longer free-streaming stages smooth out these inhomogeneities, restoring the universal spectral behavior characteristic of a nearly perfect fluid.
\section{Scaling the Transverse-Mass (\(m_T\)) Spectra}
\label{sec:transverse_mass_spectra}
\begin{figure}[ht]
  \centering
  \includegraphics[width=\linewidth]{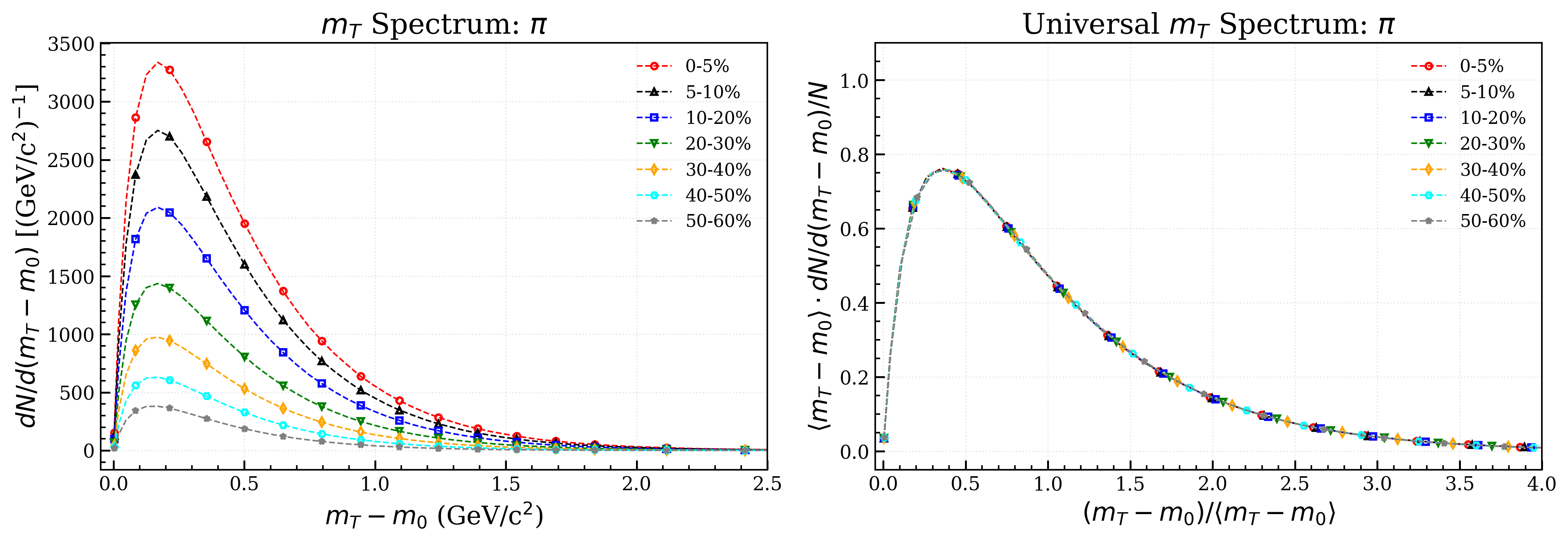}
  \caption{Scaled transverse-mass spectra for pions in Pb--Pb collisions at $\sqrt{s_{NN}}=2.76$ TeV obtained from hydrodynamic simulations using the CE MAP parameters. \textbf{Left:} transverse-mass spectra for several centrality classes. \textbf{Right:} scaled spectra $U_{m_T}(x_{m_T})$ defined in Eq.~\ref{eq:scaled_mass_spectra}, showing an approximate universality across centralities.}
  \label{fig:mass_spectra_pions}
\end{figure}

\begin{figure}[ht]
  \centering
  \includegraphics[width=\linewidth]{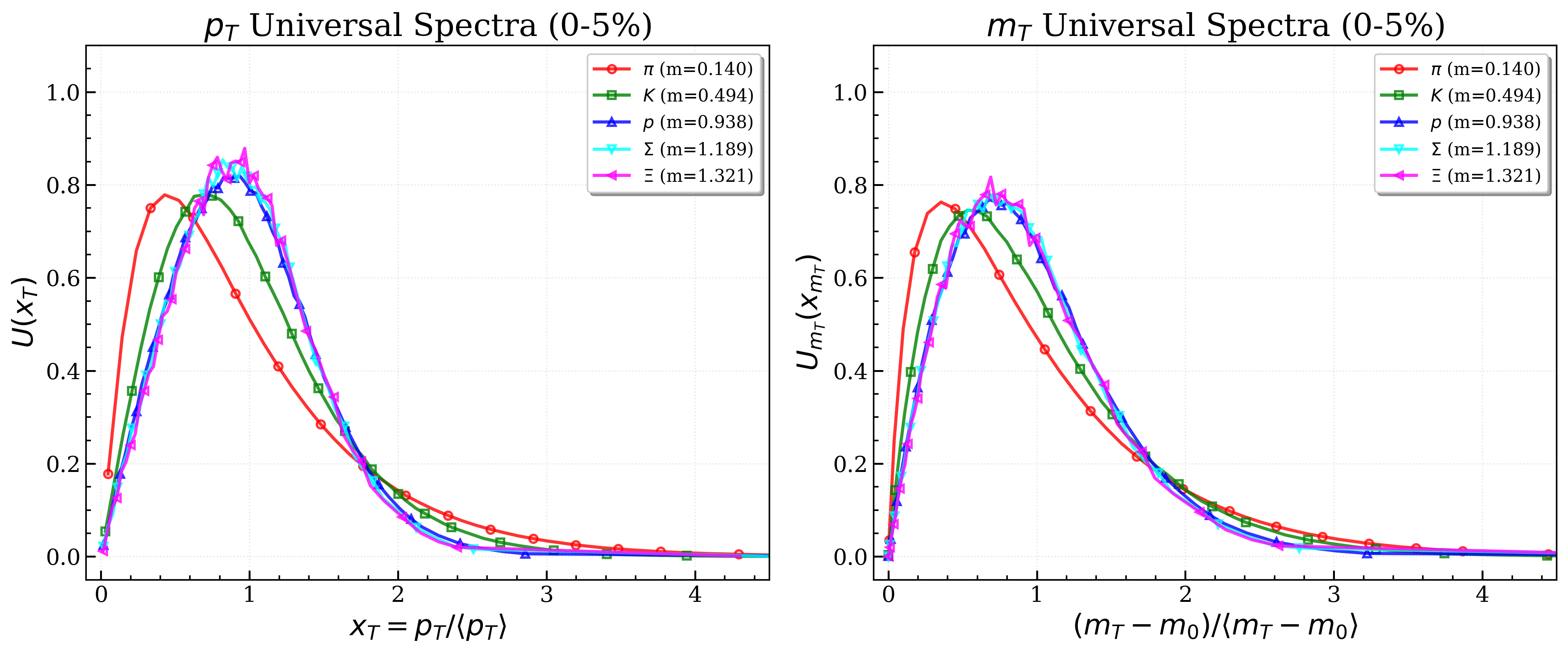}
  \caption{Comparison of scaled spectra for different hadron species ($\pi$, $K$, $p$, $\Sigma$, $\Xi$) in the most central Pb--Pb collisions. \textbf{Left:} scaled transverse-momentum spectra $U(x_T)$. \textbf{Right:} scaled transverse-mass spectra Reduced transverse-mass spectra for the same hadron species $U_{m_T}(x_{m_T})$. The transverse-mass scaling shows a slightly stronger universality across hadron species.}

  \label{fig:unviersality_pt_mass}
\end{figure}

A compelling feature of relativistic heavy-ion collisions is that 
single-particle transverse-mass spectra, \(\mathrm{d}N/\mathrm{d}m_T\),  measured across different particle species, centrality classes, and collision systems, approximately collapse onto a single ``universal'' curve upon suitable rescaling~\cite{Schaffner-Bielich:2001dlb}. The transverse mass \(m_T\) is defined as
\begin{align}
m_T &\equiv \sqrt{p_T^2 + m_0^2}, \label{eq:m_T_def}
\end{align}
where $m_0$ is the invariant rest mass of a particle, and $p_T$ is its transverse momentum. The quantity \(m_T - m_0\) is then the 
transverse kinetic energy, which provides a species-independent kinematic variable once the rest-mass offset is subtracted. Single-particle spectra can be transformed into the single-particle transverse-mass spectra using the Jacobian,
\begin{equation}
  \frac{\mathrm{d}N}{\mathrm{d}m_T}
  = \frac{\partial p_T}{\partial m_T}\,\frac{\mathrm{d}N}{\mathrm{d}p_T}
  = \frac{p_T}{m_T}\,\frac{\mathrm{d}N}{\mathrm{d}p_T},
  \label{eq:Jacobian}
\end{equation}
where the second equality follows directly from differentiating Eq.~\eqref{eq:m_T_def}, giving \(\partial m_T/\partial p_T = p_T/m_T\). Motivated by the scaling of transverse spectra in Eq.~\ref{eq:universal}, we can define a scaled mass spectrum
\begin{equation}
U_{m_T}(x_{m_T}) \equiv \frac{\media{m_T - m_0}}{N}\frac{dN}{d(m_T - m_0)}=\frac{1}{N}\frac{dN}{dx_{m_T}}     
\label{eq:scaled_mass_spectra}
\end{equation}
where $x_{m_T} = (m_T - m_0)/\media{m_T - m_0}$ as shown in Fig.~\ref{fig:mass_spectra_pions} for pions in Pb--Pb collisions at \(\sqrt{s_{NN}} = 2.76\)~TeV across several centrality classes. This scaling procedure shows a remarkable universality over centrality in the same sense as for $U(x_T)$. In Figure~\ref{fig:unviersality_pt_mass}, we display the scaled mass-spectra comparing different hadron species, showing slightly better universality over species than the scaled spectra $U(x_T)$.  
\section{Conclusions and discussion}
\label{sec:conclusions}
In this work, we investigated the universality properties of scaled transverse-momentum spectra in ultrarelativistic heavy-ion collisions. By removing the global scales associated with multiplicity and mean transverse momentum, the observable $U(x_T)$ isolates the intrinsic spectral shape and reveals an approximate universality across collision centralities and systems. Hydrodynamic simulations reproduce this behavior on an event-by-event basis, reinforcing the interpretation that the scaling may originate from the collective expansion of the QGP.

Incorporating the scaled spectra into a Bayesian framework demonstrates that this observable carries independent constraining power on the parameters of hybrid hydrodynamic models. In particular, the spectral shape is found to be sensitive to the free-streaming time, the magnitude and temperature dependence of the bulk viscosity, and the granularity of the initial state characterized by the nucleon width $w$. The comparison with the original JETSCAPE calibration reveals a noticeable tension between the parameter regions preferred by scaled spectra and those favored by $p_T$-integrated observables.

A systematic analysis of deviations from universality across the parameter space shows that the breakdown of scaling is primarily driven by initial-state granularity and pre-equilibrium dynamics rather than by uncertainties in viscous transport coefficients. These results indicate that the near-universal spectral shape emerges from a balance between collective smoothing during hydrodynamic evolution and fluctuations inherited from the initial state.

We also explored an analogous scaling for transverse-mass spectra and observed a similar universality across centralities, with indications of an even stronger universality across hadron species. These findings suggest that scaled spectra provide a complementary probe of collective dynamics and can help constrain the microscopic mechanisms governing particle production.  
\section*{Acknowledgements}
We thank the JETSCAPE Collaboration for providing access to their hybrid simulation framework, Bayesian analysis tools, and simulation data used in this work. We also acknowledge the ALICE Collaboration for making their experimental data publicly available. T.S.D. acknowledges financial support from the Brazilian National Council for Scientific and Technological Development (CNPq) through a PhD fellowship (Proc. No.\ 141432/2025-0), and from the INCT-FNA research project (Proc. No.\ 464898/2014-5, and from the Brazilian Federal Agency for Support and Evaluation of Graduate Education (CAPES), Brazil, under grant No.\ 88881.220538/2025-01. T.S.D. also acknowledges support from FAPERJ for conference accommodation through an event support grant.  M.L. was supported by FAPESP projects 2017/05685-2 and 2018/24720-6, and by project INCT-FNA Proc. No. 464898/2014-5.  
\section*{Data Availability}
All the data that support the findings of this proceedings are openly available on this GitHub repository~\cite{DominguesUniversalityRepo2025}.

\end{document}